\documentclass[journal]{IEEEtran}
\ifCLASSINFOpdf
\else
\fi
\usepackage{amsmath}
\usepackage{subfigure}
\usepackage{graphicx}
\usepackage{multicol}
\usepackage{float}
\usepackage{mathrsfs}
\makeatletter

\newcommand{\Rmnum}[1]{\expandafter\@slowromancap\romannumeral #1@}
\makeatother
\usepackage{multirow}
\usepackage{multicol}
\usepackage{adjustbox}
\usepackage{graphicx}
\usepackage{color}
\renewcommand{\baselinestretch}{1}
\usepackage[normalem]{ulem}
\usepackage{graphics}
\usepackage{ulem}
\usepackage{multirow}
\usepackage{ulem}

\newfont{\myscript}{cmmib5  scaled\magstep4}

\usepackage{slashbox}
\usepackage{amsmath,amssymb,amsfonts}\usepackage{graphicx}
\usepackage{epsfig}
\usepackage{color}
\usepackage{makecell}
\usepackage{cite}
\usepackage{multirow}

\begin{document}
%
\title{Pulse Index Modulation}

%

\author{Sultan~Aldirmaz-Colak,~\IEEEmembership{Member,~IEEE,}
        Erdogan~Aydin,
        Yasin~Celik,
        Yusuf~Acar,
        and~Ertugrul~Basar,~\IEEEmembership{Senior Member,~IEEE}
        ~
\thanks{S. Ald\i rmaz-\c{C}olak is with the Department of Electronics and Communication Engineering, Kocaeli University, Kocaeli 41380, Turkey (e-mail:
	sultan.aldirmaz@kocaeli.edu.tr)}
\thanks{E. Aydın is with the Department of Electrical and Electronics
	Engineering, Istanbul Medeniyet University, Istanbul 34857, Turkey (e-mail:
	erdogan.aydin@medeniyet.edu.tr)}
\thanks{Y. Celik is with the Department of Electrical and Electronics
	Engineering, Aksaray University, Aksaray 68100, Turkey (e-mail:
	yasincelik@aksaray.edu.tr)}
\thanks{Y. Acar is with the Group of Electronic Warfare Systems,
	STM Defense Technologies Engineering, Inc., 06510,  Ankara, Turkey (e-mail: yusuf.acar@stm.com.tr)}
\thanks{E. Basar is with the CoreLab, Department of Electrical and
	Electronics Engineering, Koç University, Istanbul 34450, Turkey (e-mail:
	ebasar@ku.edu.tr)}
\thanks{Manuscript received April 19, 2021; revised August 26, 2021.}
}

\maketitle
\begin{abstract}

Emerging systems such as Internet-of-things (IoT) and machine-to-machine (M2M) communications have strict requirements on the power consumption of used equipments and associated complexity in the transceiver design. As a result, multiple-input multiple-output (MIMO) solutions might not be directly suitable for these system due to their high complexity, inter-antenna synchronization (IAS) requirement, and high inter-antenna interference (IAI) problems. In order to overcome these problems, we propose two novel index modulation (IM) schemes, namely pulse index modulation (PIM) and generalized PIM (GPIM) for single-input single-output (SISO) schemes. The proposed models use well-localized and orthogonal Hermite-Gaussian pulses for data transmission and provide high spectral efficiency owing to the Hermite-Gaussian pulse indices. Besides, it has been shown via analytical derivations and computer simulations that the proposed PIM and GPIM systems have  better error performance and considerable signal-to-noise ratio (SNR) gain compared to existing spatial modulation (SM), quadrature SM (QSM), and traditional $M$-ary systems.

\end{abstract}

\begin{IEEEkeywords}
Hermite-Gaussian pulses, Internet-of-things (IoT), index modulation (IM), machine-to-machine (M2M), single-input single-output (SISO).
\end{IEEEkeywords}

%
\IEEEpeerreviewmaketitle

\section{Introduction}
Internet-of-things (IoT) and machine-to-machine (M2M) communications are emerging technologies which are supported by 5G. According to recent reports in 2020, the number of IoT connections exceeded the number of non-IoT ones. By 2025, it is expected that there will be more than 30 billion IoT connections, thus the efficient use of the spectrum comes to the fore in the system design \cite{X}. IoT and M2M systems should have low-power equipments and low complexity. Therefore, single-input single-output (SISO) solutions are one step ahead of their multiple-input multiple-output (MIMO) counterparts.

MIMO transmission provides transmitter (Tx) and receiver (Rx) diversity gain and increases the data rate \cite{telatar1999capacity}. However, inter-antenna synchronization (IAS) and inter-antenna interference (IAI) are big problems of MIMO schemes. 
In 2008, Mesleh \textit{et al.} proposed a novel transmission scheme, namely spatial modulation (SM) to overcome the aforementioned problems of MIMO systems \cite{mesleh2008spatial}. SM has attracted quite a lot of attention from researchers. According to this technique, only one antenna is activated among all transmit antennas for transmission at one symbol duration. Incoming bits are separated into index bits, which determine the active antenna indices, and modulated bits, which constitute symbols. Hence, incoming bits are conveyed not only by symbols but also by active antenna indices.
Since only one antenna is activated at one symbol time, IAI and IAS requirement are avoided. However, it is  stated in \cite{da2017study,ishibashi2014effects} that antenna switching in each symbol duration results in decreasing spectral efficiency (SE) in practice. Moreover, only one RF chain is sufficient in the SM, still multiple antennas are needed.
To further exploit indexing mechanisms, the concept of SM has been generalized to other resources of communication systems and index modulation (IM) has emerged. For instance, Basar \textit{et al.} proposed orthogonal frequency division multiplexing index modulation (OFDM-IM) that provides not only higher SE but also improved performance compared to classical OFDM \cite{bacsar2013orthogonal}. In OFDM-IM, subcarriers are divided into groups and in each group only a few subcarriers are activated according to index bits to convey modulated symbols. To further improve SE compared to OFDM-IM, Mao \textit{et al.} proposed dual-mode OFDM-IM \cite{mao2016dual} which utilizes entire subcarriers for symbol  transmission, unlike OFDM-IM. These techniques become very popular, then their variants have been proposed in a very short time \cite{cheng2015enhanced,ccolak2018adaptive}.  However, in the recent past, IM has been utilized to select orthogonal codes in code-division multiple access (CDMA) communication \cite{kaddoum2015code}. Both OFDM and CDMA are utilized for wideband communication however, their complexities are relatively high. Particularly, SM and general MIMO schemes require the estimation of all channels between each Tx-Rx antenna. Thus, their receiver complexity and overhead of the channel estimation are quite high. Unlike these traditional techniques, in this letter we focus on novel IM-based low complexity transceiver designs with single antennas for use in applications such as IoT, M2M with high SE.

Hermite-Gaussian pulses are widely used in ultra wide-band communication \cite{aldirmaz2010spectrally, kurt2009throughput}. In \cite{aldirmaz2010spectrally}, the authors proposed a spectrum efficient communication system that uses a summation of binary phase shift-keying (BPSK) modulated Hermite-Gaussian pulses with different orders. Since these pulses are orthogonal to each other, the transmission of a linear combination of these pulses that carries different symbols provides higher SE.

In this letter, we propose two novel IM schemes that activate certain Hermite-Gaussian pulse shapes for transmission instead of antenna indices according to the incoming information bits. Unlike SM and OFDM-IM systems, proposed pulse index modulation (PIM) and generalized PIM (GPIM) do not require multiple antennas or multiple subcarriers, and the transmitters of PIM and GPIM  have a relatively low complexity. Thus they are suitable for M2M or IoT applications.

%

The main contributions of the letter are summarized as follows:
\begin{itemize}
	\item We introduce two novel PIM schemes for SISO systems. Similar to SM, index bits are used as an extra dimension to convey data bits besides conventional constellation mapping. 
	\item We propose a low complexity detector that requires only one Tx-Rx channel state information (CSI) estimation. Since the overhead of channel estimation is low, more data can be transmitted during the coherence time.
	\item To increase the SE of the PIM technique, more than one pulse can be sent together owing to orthogonality property of Hermite-Gaussian pulses. Therefore, we introduce the GPIM scheme for SISO systems.
    \item We also obtain the average bit error probability (ABEP) for the maximum likelihood (ML) detector. ABEP results match well with the simulation results.

\end{itemize}

The remainder of this letter is organized as follows. In Section II, we introduce the system model of PIM and GPIM schemes. In Section III, performance analysis of the proposed two schemes are presented. Simulation results and performance comparisons are given in Section IV. Finally, the letter is concluded in Section V.

\textit{\textbf{Notation}}: Throughout the letter, scalar values are italicized, vectors/matrices are presented by bold lower/upper case symbols. The transpose and the conjugate transpose are denoted by $(\cdot)^T$ and $(\cdot)^H$. $\lfloor.\rfloor$, $||.||$, and $C(. , .)$ represent the floor operation, Euclidean norm, and Binomial coefficient, respectively. $\mathcal{CN}(0, \sigma^2)\ $ represents the complex Gaussian distribution with zero mean and variance $\sigma^2$ and $\textbf{I}_n$ is the  $n \times n$ identity matrix. Last, $\frac{\partial }{\partial  t}$ represents partial derivative.



\begin{figure}[t]
	\centering
	\epsfxsize 3.6in
	\leavevmode
	\epsffile{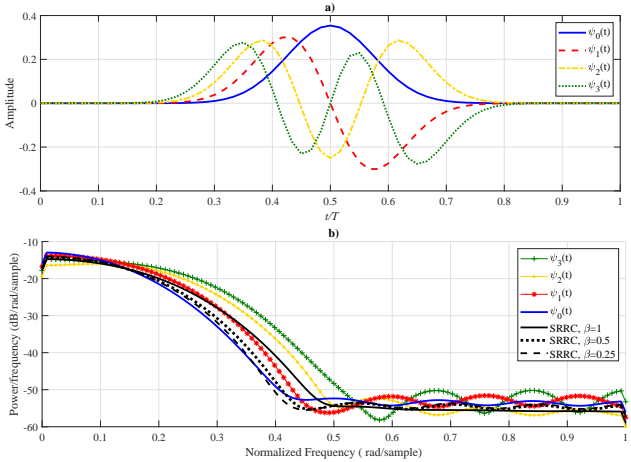}
	\vspace{-0.05cm}
	\caption{Different  Hermite-Gaussian pulses (a) with  $L = 127$ in time domain (b) bandwidth comparison with SRRC in frequency domain.}
	\label{time_bandwidth}
	\vspace{-0.5cm}
\end{figure}

\vspace{-0.3cm}

\section{System Model of Pulse Index Modulation}

We 	constitute a set of Hermite-Gaussian functions $\psi_v(t) $ that span the Hilbert space. These functions are known for their ability to be highly localized in time and frequency domains. They are defined by a Hermite polynomial modulated with a Gaussian function as

\begin{equation}
\psi_v(t) = \frac{2^{1/4}}{\sqrt{2^v v!}}H_v(\sqrt{2 \pi}t) e^{-\pi t^2},
\end{equation}
where $t$ represents time index, $v$ is the order of Hermite-Gaussian function, and  $H_v(t)$ is the Hermite polynomial series that is expressed as $H_v(t) = (-1)^v e^{t^2}\frac{\partial ^v}{\partial t^v}e^{-t^2}$.

A number of Hermite polynomials can be given for $v=0,1,2,3$ as follows:
\begin{equation}
H_0(t)\! =\! 1,\! \ H_1(t)\! =\! 2t,\! \ H_2(t)\! =\! 4t^2\!-\!2, \ H_3(t)\! =\! 8t^3\!-\!12t.
\end{equation}
One of the important properties of Hermite-Gaussian functions is orthogonality among them, which can be expressed as $\int_0^{\infty} {{\psi _m}(t)} {\psi _n}(t)dt = 	0 $ for $ m \ne n$ \cite{ozaktasfractional}.  


Representation of Hermite-Gaussian pulses  $\psi_{0}(t)$, $\psi_{1}(t)$, $\psi_{2}(t)$, and $\psi_{3}(t)$  in the time-domain are shown in Fig.~\ref{time_bandwidth}(a) for $v = 0,1,2,3$. As seen from Fig.~\ref{time_bandwidth}(a), as the order of Hermite-Gaussian pulse increases, the oscillation of the pulse also increases.
The frequency domain representation of the these Hermite-Gaussian pulses and square root raised cosine (SRRC) pulse with different roll-off factors ($\beta$) are given in Fig.~\ref{time_bandwidth}(b). We can state two important issues from Fig.~\ref{time_bandwidth}(b). First, the bandwidth of the Hermite-Gaussian pulses increases with the increase of order, as the first-null bandwidth is considered. Second, the bandwidth of the zeroth and the first order Hermite-Gaussian pulses are narrower than that of SRRC, while the bandwidth of the second and the third order Hermite-Gaussian pulses are wider. In other words, as can be seen from Fig.~\ref{time_bandwidth}(b), the bandwidth usage of the proposed scheme is relatively higher than the SRRC with $\beta= 1$ .


\begin{figure*}[!t]
	\centering
	\epsfxsize 7in
	\leavevmode
	\epsffile{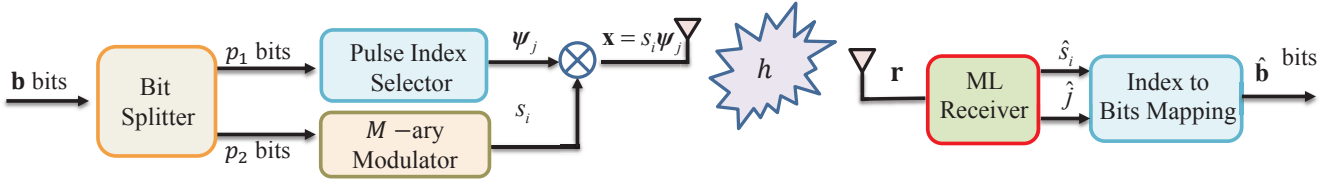}
	\vspace{-0.2cm}
	\caption{The transceiver block diagram  of the PIM scheme for $k = 1$.}
	\label{fig:Tx_Rx1}
	\vspace{-0.5cm}
\end{figure*}

\begin{figure*}[!t]
		\centering
	\epsfxsize 7in
	\leavevmode
	\epsffile{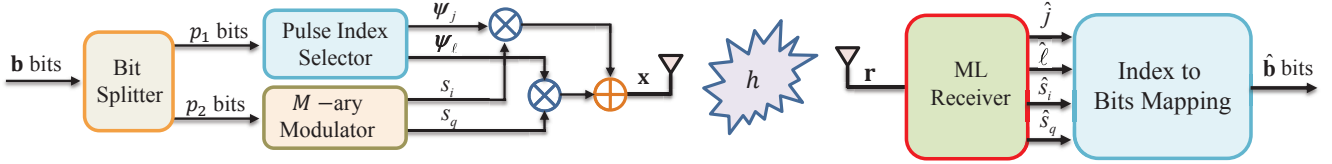}
	\caption{The transceiver block diagram of the GPIM scheme for $k = 2$.}
	\label{fig:sistem2}
\end{figure*}

In the following sections, to simplify  presentation, only discrete signal samples will be used. Discrete representation of the Hermite-Gaussian pulses are obtained from the continuous Hermite-Gaussian pulses by using Nyquist sampling theorem ($\psi_j(t) = \psi_j [lT_s]$, where $ l $ is an integer ($l=0,1,\dots L$, where $L$ is number of samples of the $j^{th}$ Hermite-Gaussian pulse) and $ T_s $ denotes the sampling interval).  Thus, each pulse is represented by a vector with  \textit{L}  samples, such as  $ \mbox{\boldmath$\psi$}_j =[\psi_{j,1} \psi_{j,2} \dots \psi_{j,L}]^T $.

	\vspace{-0.4cm}
\subsection{PIM Transmission Model ($k = 1$)}
\label{sec:Three_1}
The transceiver block diagram of the proposed PIM scheme for $k = 1$ is represented in Fig.~\ref{fig:Tx_Rx1} (at the top of the next page), where $k$ is the number of selected pulses. Firstly, incoming bit sequence \textbf{b} with the size of $\big(1 \times p_{\text{PIM}} \big)$   is splitted into pulse index selector and $M$-ary signal constellation blocks. While the first $p_1=\left \lfloor{\log_2{n \choose k}}\right \rfloor $ bits determine the active pulse shape, where $n$ denotes to total number of pulses, and the remaining  $p_2 = \log_2(M)$ bits determine the modulated symbol according to modulation scheme, where \textit{M} denotes the modulation order. In the PIM model, only one pulse is active for transmission. Thus, the baseband  PIM-based  pulse signal to be transmitted can be expressed as
\begin{equation}
\textbf{x} = s_i \mbox{\boldmath$\psi$}_j,
\end{equation}
where $s_i$ and $\mbox{\boldmath$\psi$}_j$ represent modulated $i^{th}$ symbol and $j^{th}$  selected Hermite-Gaussian pulse vector, respectively, and $i\in \{1,2,\ldots,M\}$ and  $j \in \{1,2,\ldots,2^{p_1}\}$.  For example, analytical expressions of the possible transmitted symbols for BPSK modulation in the time domain are given in Table \ref{tab1} with $s_i=\pm1$. The number of transmitted bits per channel use (bpcu) can be calculated as $p_{\text{PIM}}=p_1 + p_2$. An example of the pulse index mapping rule for  $p_1 = 2$ bits is given  by Table  \ref{tab2}. In this case, the PIM scheme for BPSK ($n = 4$, $k = 1$,  and $p_1$ = 2) transmits $p_{\text{PIM}}=p_1+p_2= 3$ bits.


%

%


	\vspace{-0.4cm}

\subsection{GPIM  Transmission Model ($k \geq 2$)}
\label{sec:Three}

To increase the transmitted bpcu of the aforementioned method, we generalized it for $k \geq 2$. In this letter, to simplify the analysis, we assume $k = 2$. The transmitter block diagram of this scheme and a look-up table, which maps the index bits to transmitted pulses, are given in Fig.~\ref{fig:sistem2} and Table \ref{tab2}, respectively.  Similar to the first method, incoming bit sequence \textbf{b} with the size of $\big(1 \times p_{\text{GPIM}} \big)$ is splitted into pulse index selector and $M$-ary signal constellation blocks. While the first $p_1 = \left \lfloor{\log_2{n \choose k}}\right \rfloor $ bits determine the active pulse shapes as given in last column of Table \ref{tab2} ($ k = 2$), the remaining $p_2 = k\log_2(M)$ bits determine the modulated symbol according to modulation scheme so that $p_{\text{GPIM}} = p_1 +p_2$. In Table \ref{tab2}, we assume that $n = 4$, $k = 2$ for the GPIM scheme. However, there are $C (4, 2)$ possible pulse shape pairs and we only use four out of them.

\begin{table}[!t]
	\centering
	\caption{Possible Pulses and Their Analytical Expressions for PIM.}
	{\begin{tabular}{|c|c|}\hline
			Possible Pulses for BPSK & Analytical Expressions\\ \hline
			$\pm \psi_0(t) $ & $ \pm 2^{1/4} e^{-\pi t^2}  $  \\ \hline
			$\pm \psi_1(t)$ & $ \pm 2^{1/4}(2 \sqrt{\pi})t e^{-\pi t^2}$\\ \hline
			$\pm \psi_2(t) $ & $ \pm \frac{2^{1/4}}{2\sqrt{2}} (8 \pi t^2 - 2) e^{- \pi t^2}$\\ \hline
			$\pm \psi_3(t)$ & $ \pm 2^{1/4}\left( \frac{4 \pi \sqrt{2 \pi} t^3 - 3\sqrt{2 \pi}t }{\sqrt{3}} \right) e^{-\pi t^2}$\\ \hline
	\end{tabular}}{}
	\label{tab1}
	\vspace{-0.3cm}
\end{table}

\begin{table}[!t]
	\centering
	\caption{A reference look-up table for $n = 4$; $k \in \{1, 2\}$, and $p_1$ = 2.}

	{\begin{tabular}{|c|c|c|}\hline

			Index Bits & \begin{tabular}[c]{@{}l@{}}Selected Pulse for\\ PIM scheme ($k = 1$) \end{tabular} & \begin{tabular}[c]{@{}l@{}}Selected Pulse for Generalized\\  PIM scheme ($k = 2$) \end{tabular} \\ \hline
			$\{0,0\}$ & $\mbox{\boldmath$\psi$}_0$  & $ \{ \mbox{\boldmath$\psi$}_0 ,\mbox{\boldmath$\psi$}_1 \}$ \\ \hline
			$\{0,1\}$ & $\mbox{\boldmath$\psi$}_1$  & $\{ \mbox{\boldmath$\psi$}_0 ,\mbox{\boldmath$\psi$}_2 \}$ \\ \hline
			$\{1,0\}$ & $\mbox{\boldmath$\psi$}_2$  & $\{ \mbox{\boldmath$\psi$}_1 ,\mbox{\boldmath$\psi$}_2 \}$ \\ \hline
			$\{1,1\}$ & $\mbox{\boldmath$\psi$}_3$  & $\{ \mbox{\boldmath$\psi$}_1 ,\mbox{\boldmath$\psi$}_3 \}$ \\ \hline

			%
	\end{tabular}}{}
	\label{tab2}
		\vspace{-0.4cm}
\end{table}

Then, the baseband  GPIM based  pulse signal to be transmitted is expressed as
\begin{equation}
\textbf{x} = \frac{1}{\sqrt{k}} (s_i \mbox{\boldmath$\psi$}_j + s_{q} \mbox{\boldmath$\psi$}_\ell),
\end{equation}
where $ \frac{1}{\sqrt{k}} $ is a normalization coefficient used to make the total symbol  energy $E_s=1$. $s_i$ and $s_q$ denote $i^{th}$ and $q^{th}$  modulated symbols respectively, $\mbox{\boldmath$\psi$}_j $ and $ \mbox{\boldmath$\psi$}_\ell $ represent the
selected Hermite-Gaussian pulses according to index bits, respectively, where $i,q \in \{1,2,\ldots,M\}$ and  $j,\ell \in \{1,2,\ldots,2^{p_1}\}$.   For example, if the information bit block is given as $ \textbf{b} = [0 \ 0 \ 1 \ 0 ] $, the selected Hermite-Gaussian pulses are zeroth order and the first order ones according to the first two bits, and selected BPSK symbols for the third and the fourth bits are  $s_i =1$ and $s_{q}=-1$. Thus the transmitted signal can be expressed as $\textbf{x} = \frac{1}{\sqrt{2}} \big( \mbox{\boldmath$\psi$}_0 - \mbox{\boldmath$\psi$}_1 \big)$, if BPSK modulation is applied. As Hermite-Gaussian pulses are orthogonal to each other, the modulation type can be thought as quadrature phase shift-keying (QPSK).

%
%
%

\begin{table}[!t]
	\centering
	\caption{Possible Pulses and Their Analytical Expressions for GPIM.}

	{\begin{tabular}{|c|c|}\hline
			Possible Pulses & Analytical Expressions\\ \hline
			$\pm \psi_0(t) \pm  \psi_1(t)$ & $2^ {1/4} (1 \pm 2 \sqrt{\pi}t) e^{-\pi t^2}$  \\ \hline
			$\pm \psi_0(t) \pm  \psi_2(t)$ & $2^ {1/4} (1 \pm  \frac {4\pi t^2 - 1}{\sqrt{2}}) e^{-\pi t^2}$\\ \hline
			$\pm \psi_1(t) \pm  \psi_2(t)$ & $2^ {1/4} (2 \sqrt{\pi} t \pm  \frac {4\pi t^2 - 1}{\sqrt{2}} ) e^{-\pi t^2}$\\ \hline
			$\pm \psi_1(t) \pm  \psi_3(t)$ & $2^ {1/4} (2 \sqrt{\pi}t \pm  \frac {4\pi \sqrt{2 \pi} t^3 - 3\sqrt(2 \pi) t}{\sqrt{3}}) e^{-\pi t^2}$\\ \hline

	\end{tabular}}{}
	\label{tab3}
		\vspace{-0.3cm}
\end{table}

Analytical expression of the possible transmitted pulses for BPSK modulation in the time domain is given in Table \ref{tab3} for GPIM.
SE of GPIM scheme can be calculated as $p_{\text{GPIM}}=\left \lfloor{\log_2{n \choose k}}\right \rfloor $ + $k \log_2(M)$ bpcu. Without losing generality, we use four different Hermite-Gaussian functions of orders $v = 0, 1, 2$, and $3$ for practical considerations and for simplicity ($n=4$, $M=2$ ). Thus, the SE of GPIM scheme equals to $p_{\text{GPIM}}=4$ bpcu while the classical SISO system with $M=2$  obtains $1$ bpcu.

\subsubsection{ML Detection of PIM and GPIM Schemes}

The  vector representation of the received baseband signal for PIM and GPIM schemes can be expressed, respectively  as follows:
\begin{eqnarray}
\mathbf{r}_{\text{PIM}} &=&  \mathbf{x}h  + \mathbf{n} \nonumber \\
&=& s_i \mbox{\boldmath$\psi$}_j  h + \mathbf{n},\\
	\mathbf{r}_{\text{GPIM}}  &=&  \mathbf{x}h  + \mathbf{n} \nonumber \\
	&=&  \big(s_i \mbox{\boldmath$\psi$}_j + s_{q} \mbox{\boldmath$\psi$}_\ell\big)h  + \mathbf{n},
\end{eqnarray}
where $\textbf{n} \in \mathcal{C}^{L\times 1}$ is the noise vector with elements following
$\mathcal{CN}(0, \frac{N_0}{2} \textbf{I}_L)\ $ and $h$ represents the complex Rayleigh fading coefficient.

The ML detectors for PIM and GPIM schemes can be expressed respectively as follows:
\begin{eqnarray}\label{eq3_2}
	\Big(\hat{s}_i,\hat j\Big)  &\!\!\!\!\!=&\!\! \!\!\! \underset{i,j}{\mathrm{arg\,max}}  \Big(\mathrm{Pr}\big(\textbf{r}_{\text{PIM}}|s_i , \mbox{\boldmath$\psi$}_j \big)\Big) \nonumber \\
	&\!\!\!\!\!=&\!\!\!\!\!   \underset{i,j}{\mathrm{arg\,min}}  \bigg\{ \Big| \Big|  \textbf{r}_{\text{PIM}} - h s_i  \mbox{\boldmath$\psi$}_j  \Big| \Big|^2 \bigg\}, \\
	\Big(\hat{s}_i,\hat{s}_q,\hat j, \hat \ell\Big) &\!\!\!\!\!=&\!\!\!\!\!  \underset{i,j,q,\ell}{\mathrm{arg\,max}}  \Big(\mathrm{Pr}\big(\textbf{r}_{\text{GPIM}}|s_i , s_q \mbox{\boldmath$\psi$}_j, \mbox{\boldmath$\psi$}_\ell \big)\Big) \nonumber \\
&\!\!\!\!\!=&\!\!\! \!\! \underset{i,j,q,\ell}{\mathrm{arg\,min}}  \bigg\{\! \Big| \Big|  \textbf{r}_{\text{GPIM}} \!-\! \big(s_i \mbox{\boldmath$\psi$}_j + s_{q} \mbox{\boldmath$\psi$}_\ell\big)h  \Big| \Big|^2 \!\bigg\}
\end{eqnarray}
where, $i,q \in \{1,2,\ldots,M\}$ and  $j,\ell \in \{1,2,\ldots,2^{p_1}\}$.

Finally, using the detected $(\hat{s}_i,\hat{s}_q,\hat j, \hat \ell)$ values, the originally transmitted bit sequence $ \hat{\mathbf{b}} $ is reconstructed at the receiver with the help of the index to bits mapping technique  as shown at the receiver block of the PIM and GPIM systems.
%
%
%

	\vspace{-0.4cm}
\section{Performance Analysis}

In this section, we analyze the ABEP performance of the  PIM and GPIM schemes. Accordingly, using the well-known union bounding technique as in \cite{Proakis}, the expression of  ABEP  $ \mathbb{P} $ for proposed two schemes can be given as follows:
\begin{equation}\label{eq24}
\mathbb{P} \le \frac{1}{2^{\tilde{p}}}\sum_{d=1}^{2^{\tilde{p}}}\sum_{z=1}^{2^{\tilde{p}}}\frac{\mathbb{P}_{\text{e}}\Big(\mbox{\boldmath$\xi$}_d\to \hat{\mbox{\boldmath$\xi$}}_z\Big)N(d,z)}{\tilde{p}},
\end{equation}
where $ \tilde{p}=p_1+p_2 $ is the number of bits transmitted in active pulse indices and modulated symbols,  $ N(d,z) $ is expressed as the number of bits in errors   between the vectors $\mbox{\boldmath$\xi$}_d$ and $\hat{\mbox{\boldmath$\xi$}}_z$. $\mathbb{P}_{\text{e}}\big(\mbox{\boldmath$\xi$}_d \to \hat{\mbox{\boldmath$\xi$}}_z\big) $ is the APEP of deciding $\hat{\mbox{\boldmath$\xi$}}_z$ giving that $\mbox{\boldmath$\xi$}_d$  is transmitted and it can be expressed as
\begin{equation}\label{eq25}
\mathbb{P}_{\text{e}}\Big(\mbox{\boldmath$\xi$}_d \to \hat{\mbox{\boldmath$\xi$}}_z\Big) =\frac{1}{2}\Bigg(1-\sqrt{\frac{\sigma^2_{k,\alpha}}{1+\sigma^2_{k,\alpha}}}\Bigg),
\end{equation}
where,  $k \in \{1,2\}$. Note that, for $k = 1$ and for  $k = 2$, the ABEPs of PIM and GPIM schemes are obtained, respectively. For the  PIM and GPIM schemes $ \sigma^2_{k,\alpha} $ is given by:
\begin{equation}\label{eq26}
\sigma^2_{1,\alpha} = \left\lbrace \begin{tabular}{c}
$\frac{E_s}{2N_0}\sigma_h^{2} \Big(|{s}_{i}|^2+|{\hat{s}}_{{i}}|^2\Big)$
$\text{if} \ \mbox{\boldmath$\psi$}_j \not= \mbox{\boldmath$\psi$}_{\hat{j}}$ \\
$\frac{E_s}{2N_0}\sigma_h^{2} \Big(|s_{{i}}-{\hat{s}}_{{i}}|^2\Big)$
$\text{if} \  \mbox{\boldmath$\psi$}_j= \mbox{\boldmath$\psi$}_{\hat{j}}$ \\
\end{tabular} \right.
\end{equation}
\begin{equation}\label{eq27}
	\!\!\sigma^2_{2,\alpha} \!\!=\!\! \left\lbrace \begin{tabular}{c}
		$\!\!\!\!\frac{E_s}{2N_0}\sigma_h^{2} \big(|{s}_{i}|^2\!+\!|{\hat{s}}_{{i}}|^2\!+\!|{s}_{q}|^2\!+\!|{\hat{s}}_{{q}}|^2\big)$
		$\text{if} \ \mbox{\boldmath$\psi$}\! \not=\! \mbox{\boldmath$\psi$}_{\hat{j}}, \mbox{\boldmath$\psi$}_\ell \not=\! \mbox{\boldmath$\psi$}_{\hat{\ell}}$ \\
		$\!\!\!\!\!\!\!\!\!\!\frac{E_s}{2N_0}\sigma_h^{2} \big(|{s}_{i}\!-\!{\hat{s}}_{{i}}|^2\!+\!|{s}_{q}|^2\!+\!|{\hat{s}}_{{q}}|^2\big)$
		$\text{if} \ \mbox{\boldmath$\psi$}_j \!=\! \mbox{\boldmath$\psi$}_{\hat{j}}, \mbox{\boldmath$\psi$}_\ell \!\not=\! \mbox{\boldmath$\psi$}_{\hat{\ell}}$ \\
		$\!\!\!\!\!\!\!\!\!\!\!\!\frac{E_s}{2N_0}\sigma_h^{2} \big(|{s}_{i}|^2\!+\!|{\hat{s}}_{{i}}|^2\!+\!|{s}_{q}\!-\!{\hat{s}}_{{q}}|^2\big)$
		$\text{if} \ \mbox{\boldmath$\psi$}_j \!\not=\! \mbox{\boldmath$\psi$}_{\hat{j}}, \mbox{\boldmath$\psi$}_\ell \!=\! \mbox{\boldmath$\psi$}_{\hat{\ell}}$ \\
		$\!\!\!\!\!\!\!\!\!\!\!\!\!\!\!\!\frac{E_s}{2N_0}\sigma_h^{2} \big(|{s}_{i}\!-\!{\hat{s}}_{{i}}|^2\!+\!|{s}_{q}\!-\!{\hat{s}}_{{q}}|^2\big)$
		$\text{if} \ \mbox{\boldmath$\psi$}_j \!=\! \mbox{\boldmath$\psi$}_{\hat{j}}, \mbox{\boldmath$\psi$}_\ell \!=\! \mbox{\boldmath$\psi$}_{\hat{\ell}}$
	\end{tabular} \right.
\end{equation}
where  $\hat{s}_{{i}}$ and $\hat{s}_{{q}}$ are estimates of ${s}_{i}$ and ${s}_{q}$, respectively.  $\sigma_h^{2}$  is the variance of Rayleigh fading channel coefficient and $\sigma_h^{2}=1$.

Consequently, by substituting (\ref{eq26}) and (\ref{eq25}) into (\ref{eq24}), we obtain the ABEP for PIM system, and similarly, by substituting (\ref{eq27}) and (\ref{eq25}) into (\ref{eq24}), we obtain the ABEP for GPIM system.
\begin{figure}[!t]
	\centering
	\epsfxsize 3.5in
	\leavevmode
	\epsffile{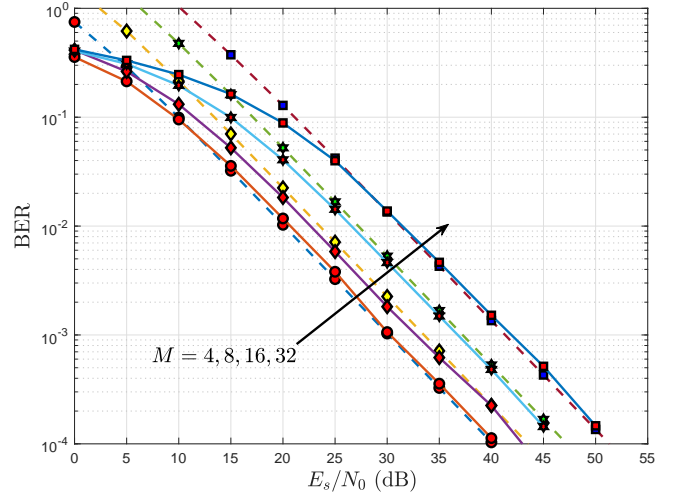}
	\vspace{-0.3cm}
	\caption{BER performance curves of the PIM scheme with PSK modulation for various $M$ values ($n = 4$, $k = 1$).}
	\label{fig:Sim3}
\end{figure}

\section{Simulation Results}
\label{section.simulation}

To demonstrate the improved performance of the proposed techniques, the bit error rate (BER) of PIM systems is evaluated with different system setups. SM,  quadrature SM (QSM), traditional $M$PSK/QAM schemes are selected as benchmarks. The SNR used in computer simulations herein is defined as $E_s/N_0$  where $E_s$ is energy per symbol and $N_0$ is the noise power. At the receiver, ML detector is used for all systems. Each Hermite-Gaussian pulse consists of $127$ samples. Since the head and tail of the blows contain a large number of zero-value samples, we truncate their edges. Thus, each pulse includes $61$ samples. All simulations are performed over frequency-flat Rayleigh fading channels. We assume that the channel is constant during one symbol duration, and the CSI is perfectly known at the receiver. 

The theoretical and simulation average BER performance  curves of the PIM scheme with $M$-PSK, $M=4,8,16, 32$, $n=4$, and $k=1$  are presented for $p_{\text{PIM}} =4,5,6$, and $7$ bits in Fig.~\ref{fig:Sim3}. Here, the PIM technique transmits $4,5,6$, and $7$ bits by $2$ bits with active pulse indices and $2,3,4$, and $5$  bits with the transmitted symbols, respectively. As can be seen from Fig.~\ref{fig:Sim3}, analytical results match simulation results well particularly at high modulation order.

\begin{figure}[!t]
	\centering
	\epsfxsize 3.4in
	\leavevmode
	\epsffile{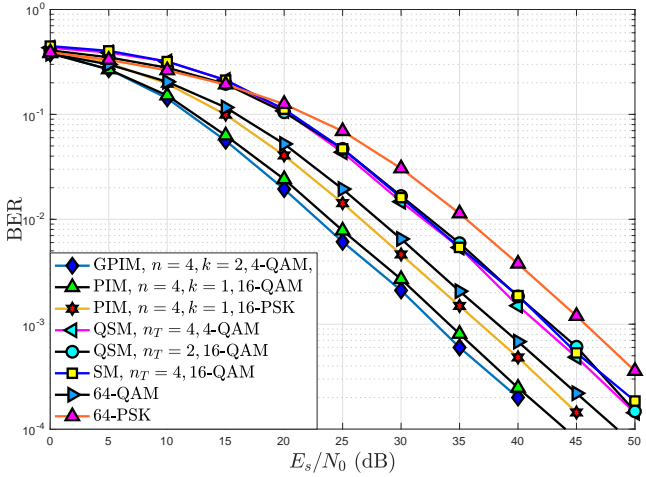}
		\vspace{-0.3cm}
	\caption{Performance comparisons of GPIM, PIM, QSM, SM, and traditional PSK/QAM systems for $6$ bpcu.}
	\label{fig:Sim1}
\end{figure}

The  average BER performance  curves of the PIM, GPIM and benchmarks schemes are shown in Fig.~\ref{fig:Sim1} for $M$-PSK/QAM at $6$ bpcu. GPIM technique carries $2$ bits with active pulse indices and $4$ bits with the transmitted symbol; PIM scheme transmits $2$ bits with with active pulse indices and $4$ bits with the transmitted symbol; the QSM technique carries either $4$ bits with antenna indices and $2$ bits with the transmitted symbol or $2$ bits with antenna indices and $4$ bits with the transmitted symbol. In the SM technique, $2$ bits are transmitted in antenna indices and $4$ bits are transmitted with symbols. In PSK/QAM, all 6 bits are carried on a modulated symbol with $M = 64$. As seen from Fig.~\ref{fig:Sim1}, the GPIM  scheme provides better performance with approximately $1$ dB SNR gain compared to PIM  system when QAM is used.  Also, the analytical and the simulation results match well. The proposed GPIM and PIM schemes have also better BER performance compared to SM, QSM, and traditional QAM schemes.

\begin{figure}[!t]
	\centering
	\epsfxsize 3.45in
	\leavevmode
	\epsffile{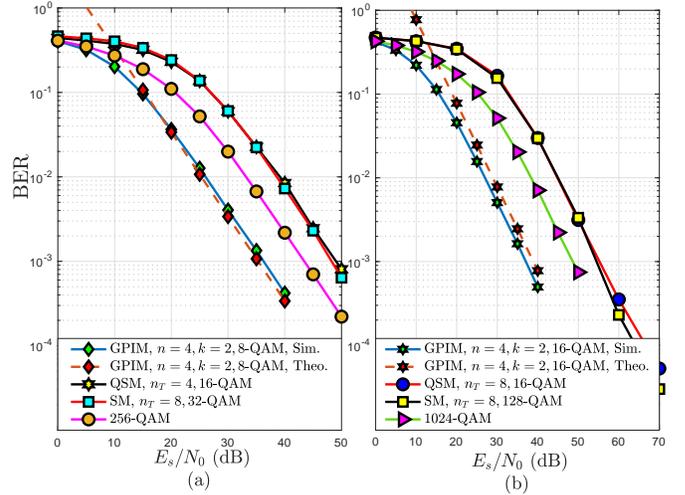}
		\vspace{-0.3cm}
	\caption{Performance comparisons of GPIM, QSM, SM,  and traditional QAM systems for (a) $8$ bpcu (b) $10$  bpcu.}
	\label{fig:Sim2}
\end{figure}

Fig.~\ref{fig:Sim2} presents average BER performance  curves of GPIM, QSM, SM, and traditional QAM  schemes  for (a) 8 bpcu and (b) 10 bpcu. For Fig.~\ref{fig:Sim2} (a), GPIM  carries  $2$ bits with active pulse indices and $6$ bits with the transmitted symbol; the QSM technique carries $4$ bits with antenna indices and $4$ bits with the transmitted symbol; the SM scheme carries $3$ bits with antenna indices and $5$ bits with the transmitted symbol.  The corresponding values in Fig. 7 (b) are $2$ and $8$ bits for GPIM;  $6$ and $4$ bits for QSM; $3$ and $7$ bits for SM schemes, respectively. In $M$-QAM, all $8$ and $10$ bits are carried on a modulated symbol with $M=256$ and $M=1024$, respectively. We can see from Fig.~\ref{fig:Sim2} that  GPIM scheme has a considerable SNR gain compared to  SM and QSM schemes for the same bpcu. At BER $= 10^{-2}$, the proposed scheme requires almost 18 dB less power compared to SM and QSM schemes for $M = 16$ case.





\section{Conclusions}
\label{section.conclusions}

We have proposed two new IM schemes, namely PIM and GPIM,  which exploit the indices of Hermite-Gaussian pulses for SISO systems. These methods are suitable for systems that need low complexity owing to their SISO structure. For this reason, we think that our schemes can be utilized especially in M2M and IoT applications. Analytical expressions for average BER of the PIM and GPIM systems have been derived and their superiority have been shown.
\ifCLASSOPTIONcaptionsoff
  \newpage
\fi

	\vspace{-0.4cm}
\bibliographystyle{IEEEtran}
\bibliography{IEEEabrv,ref2.bib}








\end{document}